# Effects of Four-Phonon Scattering and Wave-like Phonon Tunneling Effects on Thermoelectric Properties of Mg$_2$GeSe$_4$ using Machine Learning

Hao-Jen You[1,!], Yi-Ting Chiang[2,!], Arun Bansil[3,4], Hsin Lin[1,*]

[1] *Institute of Physics, Academia Sinica, Taipei 115201, Taiwan*

[2] *Institute of Statistical Science, Academia Sinica, Taipei, 115201, Taiwan*

[3] *Department of Physics, Northeastern University, Boston, MA, 02115, USA*

[4] *Quantum Materials and Sensing Institute, Northeastern University, Burlington, MA 01803, USA*

[!] *The authors contributed equally to this work*

**Corresponding author** * nilnish@gmail.com

**ABSTRACT**

We present a machine-learning interatomic potential (MLIP) framework, which substantially accelerates the prediction of lattice thermal conductivity for both particle-like and wave-like thermal transport, including three-phonon and four-phonon scattering processes, and achieves speedups of five orders of magnitude compared to the conventional DFT calculations. We illustrate our approach through an in-depth study of $Mg_2GeSe_4$ as an exemplar thermoelectric material. Four phonon scattering is found to reduce lattice thermal conductivity by 22.5% at 300K and 26.7% at 900 K. The particle-like contribution to lattice thermal conductivity decreases, while the wave-like component increases with increasing temperature. The maximum figure of merit zT at 900K is found to be 0.49 for the n-type and 0.45 for the p-type $Mg_2GeSe_4$, respectively. Our analysis reveals that a substantial contribution to the Seebeck coefficient is driven by multi-band features in the valence band. Our study gives insight into the promising thermoelectric performance of $Mg_2GeSe_4$ and provides an efficient and accurate scheme for developing rational materials discovery strategies and practical applications of thermoelectric materials.

# 1. Introduction

Chalcogenides[1] are among the most prominent materials for renewable energy applications. The efficiency of a thermoelectric material,[2–4] which can convert heat into electricity, is governed by the dimensionless figure of merit, $zT = \frac{S^2\sigma}{\kappa_e+\kappa_L}$, where $\sigma$, $S$, $T$, $\kappa_e$, and $\kappa_L$ are the electrical conductivity, Seebeck coefficient, absolute temperature, electrical thermal conductivity, and lattice thermal conductivity, respectively. An ideal thermoelectric would thus possess high electrical conductivity like the metals, a large Seebeck coefficient like the semiconductors, and an ultralow thermal conductivity like the glasses[5]. Achieving these competing thermal and electrical characteristics in a material presents a challenge. Among the chalcogenides, PbTe[6,7], GeTe[8,9], and BiCuTeO[10,11] are the top performers in the mid-temperature range (600–900 K) due to their superior thermoelectric properties, although the toxicity of lead (Pb) restricts widespread use of lead based chalcogenides[12,13], necessitating the development of lead-free alternatives using earth-abundant elements[14,15].

Mg$_2$GeSe$_4$ is a chalcogenide, which has been synthesized using a solid-state reaction method, and its optical properties have been investigated[16]. Mg$_2$GeSe$_4$ exhibits stability in air, which is important for developing applications[17–19]. Here we discuss the potential of Mg$_2$GeSe$_4$ as a promising thermoelectric material using a highly efficient computational scheme based on machine-learning interatomic potentials (MLIPs) combined with message-passing neural networks (MPNNs).

Most MLIPs are based on Behler's framework[20], where the total energy is calculated by summing the atomic energies, each influenced by its local atomic environments within a cutoff radius. This setup enables scalable modeling and employs descriptors such as ACSF[20–22], NEP[23–26], SOAP[27], and DeePMD[28] to ensure rotation and translation invariance. MPNNs[29] offer an alternative by modeling atoms and bonds as nodes and edges in a graph, allowing information exchange beyond the cutoff radius and enabling the model to capture more complex interactions between atoms. While traditional descriptors are widely used in thermoelectric material calculations[30–33], they face limitations in completeness, higher-order interactions, and the need for extensive data. MLIPs based on MPNNs can achieve high accuracy using approximately an order of magnitude less training data, reducing the computational cost associated with generating data from first-principles calculations. This makes MPNNs a promising approach for thermoelectric property calculations[34–36].

A recent unified theory of thermal transport identifies two primary mechanisms: particle-like and wave-like transport.[37,38] In ordered crystalline materials, particle-like thermal transport is the dominant mechanism; however, wave-like thermal transport cannot be ignored, particularly in materials with large unit cells. The unit cell of $Mg_2GeSe_4$ consists of 28 atoms, resulting in phonon band structures with numerous phonon branches of narrow bandwidth. Consequently, wave-like tunneling between different branches may contribute significantly to thermal transport. Moreover, to comprehensively capture the full scope of thermal transport, it is essential to account for both three-phonon and four-phonon scattering processes. This requirement poses a significant challenge for conventional Density Functional Theory (DFT) calculations. By combining the accuracy of DFT and the computational efficiency of MPNNs, we accurately delineate temperature-dependent contributions to the thermal and thermoelectric properties of both n-type and p-type $Mg_2GeSe_4$ resulting from multi-band features in the electronic structure as well as the contributions to thermal transport involving three and four phonon scattering processes in both the particle and wave-like transport regimes. Our study gives insight into the promising thermoelectric properties of $Mg_2GeSe_4$ and provides an efficient machine-learning-based approach for rational design of high-performance thermoelectric materials with tailored properties for energy applications.

## 2. Computational Methodology

2.1 Density Functional Theory Calculations

First-principles calculations were carried out within the framework of the Density Functional Theory (DFT) using the projector augmented wave (PAW)[26] method as implemented in the Vienna Ab initio Simulation Package (VASP)[40,41]. The generalized gradient approximation (GGA)[42] of Perdew–Burke–Ernzerhof (PBE)[43] for solids (PBEsol)[44] was applied for treating exchange-correlation effects. A 28 atom primitive cell was used in the computations. Structural optimization was performed without constraints on cell shape or cell size until the forces on each atom were less than $5 \times 10^{-4}$ eV/Å. A plane-wave energy cutoff of 520 eV, an energy convergence threshold of $10^{-8}$ eV, and an $8 \times 6 \times 4$ Monkhorst–Pack k-point mesh[45] were employed for structural relaxation. Recognizing that local-density approximation (LDA) and GGA tend to underestimate band gaps, electronic properties were calculated using the Heyd–Scuseria–Ernzerhof (HSE06) hybrid functional[46], which was also utilized for electronic transport calculations. Relativistic effects were incorporated by including spin-orbit coupling (SOC) effects in electronic structure calculations. The VASPKIT code[47] was used for pre- and post-processing and the crystal structure was plotted using the VESTA package[48].

2.2 Electrical Transport Calculations

The AMSET package[49] was utilized to calculate electronic transport properties, providing a more accurate alternative to the commonly used constant relaxation time approximation (CRTA), which often overestimates the figure of merit zT.[49,50] In contrast, AMSET package applies the momentum relaxation time approximation (MRTA), explicitly computing scattering rates for individual electronic states within the Born approximation. By utilizing the relaxation time approximation, the electrical transport distribution function is derived through the linearized Boltzmann transport equation (BTE): $\sigma^{\alpha\beta}(E) = \sum_i \int \frac{d\mathbf{k}}{8\pi^3} v_{i\mathbf{k}}^\alpha v_{i\mathbf{k}}^\beta \tau_{i\mathbf{k}}^{el} \delta(E - E_{i\mathbf{k}})$, where $\tau_{i\mathbf{k}}^{el}$ and $\delta$ represent the electronic relaxation time and Dirac delta function, respectively. Here, $E$, $\mathbf{k}$, and $i$ denote electron energy, wave vector, and electronic band index. AMSET package incorporates distinct electronic relaxation times for various scattering mechanisms, including acoustic deformation potential (ADP), ionized impurity (IMP), polar optical phonon (POP), and piezoelectric (PIE) scattering. Parameters essential to these calculations—such as the deformation potential, bulk modulus, static dielectric constant, effective phonon, and high-frequency dielectric constant—were obtained from first-principles calculations. A dense interpolated k-grid of 45 × 37 × 21 was used to calculate the Seebeck coefficient, electrical conductivity, and electronic thermal conductivity. Transport properties were converged with respect to the interpolation factor, as demonstrated in Fig. S1, with detailed calculation parameters provided in Table S1.

2.3 Preparing and Training MLIP

To construct MLIP, *ab initio* molecular dynamics (AIMD) simulations were conducted on supercells of varying sizes (28, 56, 112, and 224 atoms), generating configurations with random displacements at temperatures from 50 K to 2000 K and under uniform triaxial strains ranging from -3% to 3%. The MACE[35,36] framework, utilizing MPNN models, was employed to learn atomic representations from this reference data. By integrating Atomic Cluster Expansion (ACE)[51] with equivariant message passing, MACE can predict scalar, vector, and tensor properties. Through advanced message-passing methods, MACE enhances learning efficiency and reduces the required dataset size, making it highly effective for training the interatomic potential. The training dataset included 1200 configurations, with 912 (76%) used for training, 48 (4%) for validation, and 240 (20%) for testing.

Our model used a 9 Å cutoff radius for atomic environments, an angular resolution of $L_{max} = 2$ ,and two message-passing layers, providing an optimal balance between accuracy and computational cost. A batch size of 10 was used for both training and validation, and the model was trained on two NVIDIA RTX 3090 GPUs (without an NVLink bridge) using float64 precision for 500 epochs, as further error reduction plateaued beyond this point. The loss function was designed to minimize errors in energy, force, and stress, with weights of 1 for energy, 100 for forces, and 1 for virials, ensuring proper minimization across different physical quantities. The root-mean-squared errors (RMSE) on the predicted energies and force tensors on the testing data set are 0.4 meV/atom and 19.5 meV/Å, respectively.

2.4 Phonon Transport Calculations

To accurately compute the lattice thermal conductivity, it is essential to precisely determine the harmonic and anharmonic interatomic force constants (IFCs), including second-, third-, and fourth-order IFCs, which serve as key inputs for solving the Peierls-Boltzmann transport equation (PBTE). We performed molecular dynamics (MD) simulations within the canonical (NVT) ensemble using MLIP to obtain the necessary trajectory data, and the corresponding energies and forces. A 2 × 2 × 2 supercell containing 224 atoms was used, and the simulation spanned 100 ps with a time step of 1 fs. Data snapshots were taken every 0.1 ps to extract information for calculating the anharmonic force constants and atomic displacements. Subsequently, compressed sensing lattice dynamics (CSLD)[52–54] was applied to determine the higher-order force constants. This approach includes adjusting the parameter α through cross-validation and determining the anharmonic force constants using an elastic grid method. The outcomes of the cross-validation are illustrated in Fig. S2 (a). Interaction cutoffs of 5.29 Å and 4.23 Å were applied for the third and fourth order force constants, respectively. Figs. S2 (c) and S2 (d) illustrate the norm of the interatomic force constant (IFC) matrices as a function of the maximum cutoff distance. It is evident that the atomic cutoff distance for converged IFCs must exceed 5.1 Å for the third-order and 4.1 Å for the fourth-order interactions, indicating that three-atom and four-atom interactions beyond these distances are negligible. Additionally, self-consistent phonon (SCP)[55] calculations were carried out using the ALAMODE code[56–60], and the PBTE solved using the revised version of the FourPhonon package[61,62] from the ShengBTE package[63].

We employed uniform 8 × 8 × 8 and 3 × 3 × 3 q-meshes for three-phonon and four-phonon processes, respectively, with the q-point grid convergence results presented in Fig. S2 (b). Due to the high computational cost of iterative solutions for four-phonon processes, we applied the Boltzmann transport equation (BTE) iteration scheme for three-phonon scattering and used the single-mode relaxation time approximation (SMRTA) for four-phonon scattering.

According to the Wigner transport equation[37,38], derived from the Wigner phase-space formulation, the lattice thermal conductivity $\kappa_L^{\alpha\beta}$ can be decomposed into particle-like $\kappa_p^{\alpha\beta}$ and wave-like $\kappa_c^{\alpha\beta}$ components, where $\alpha$ and $\beta$ denote Cartesian directions. The particle-like phonons contribute to the thermal conductivity of the phonon population, calculated from the diagonal terms (where the $i$th mode equals the $j$th mode) of the Wigner heat-flux operator. This term, analogous to the particle-like thermal conductivity, is given by: $\kappa_p^{\alpha\beta} = \frac{1}{VN_\mathbf{q}}\sum_{\mathbf{q}i} C_\mathbf{q}^i v_{\mathbf{q},\alpha}^i v_{\mathbf{q},\beta}^i \tau_\mathbf{q}^i$. The wave-like thermal conductivity $\kappa_c^{\alpha\beta}$, arising from off-diagonal $i$th mode not equal to $j$th mode terms, captures interbranch tunneling effects between the distinct phonon modes $i$ and $j$ and it is expressed as: $\kappa_c^{\alpha\beta} = \frac{\hbar^2}{k_B T^2 V N_\mathbf{q}} \sum_\mathbf{q} \sum_{i \neq j} \frac{\omega_\mathbf{q}^i + \omega_\mathbf{q}^j}{2} v_{\mathbf{q},\alpha}^{i,j} v_{\mathbf{q},\beta}^{i,j} \times \frac{\omega_\mathbf{q}^i n_\mathbf{q}^i (n_\mathbf{q}^i+1) + \omega_\mathbf{q}^j n_\mathbf{q}^j (n_\mathbf{q}^j+1)}{4(\omega_\mathbf{q}^i - \omega_\mathbf{q}^j)^2 + (\Gamma_\mathbf{q}^i + \Gamma_\mathbf{q}^j)^2}(\Gamma_\mathbf{q}^i + \Gamma_\mathbf{q}^j)$. Here, $\hbar$ denotes the reduced Planck constant, $k_B$ the Boltzmann constant, $V$ the unit cell volume, and $N_\mathbf{q}$ the number of sampled phonon wave vectors in the first Brillouin zone. The parameters $C_\mathbf{q}^i$, $v_\mathbf{q}^i$, $\tau_\mathbf{q}^i$, $\omega_\mathbf{q}^i$, $v_\mathbf{q}^{i,j}$, and $\Gamma_\mathbf{q}^i$ (where $\tau_\mathbf{q}^i = \frac{1}{\Gamma_\mathbf{q}^i}$) refer to the heat capacity, group velocity, phonon lifetime, frequency, interbranch phonon group velocity, and phonon scattering rate, respectively. The Bose-Einstein distribution is represented by $n_\mathbf{q}^i = [\exp(\hbar\omega_\mathbf{q}^i/k_B T) - 1]^{-1}$. This decomposition provides a detailed view of the contributions to thermal conductivity, distinguishing between population-driven and coherence-related mechanisms.

## 3. RESULTS

### 3.1 Equilibrium Geometry and Electronic Structure

Mg$_2$GeSe$_4$ is a ternary chalcogenide semiconductor. The calculated lattice constants for Mg$_2$GeSe$_4$ are: a = 6.31 Å, b = 7.81 Å, and c = 13.42 Å, which are smaller than the reported experimental values[16] of: a = 6.36 Å, b = 7.85 Å, and c = 13.50 Å. Given that Mg$_2$GeSe$_4$ crystallizes in the orthorhombic *Pnma* symmetry, we calculated nine independent elastic constants to assess its mechanical properties, as detailed in Table S1. The calculated elastic constants satisfy the Born–Huang stability criteria, confirming mechanical stability: $C_{11} > 0$, $C_{11}C_{22} > C_{12}^2$, $C_{11}C_{22}C_{33} + 2C_{12}C_{13}C_{23} - C_{11}C_{23}^2 - C_{22}C_{13}^2 - C_{33}C_{12}^2 > 0$, $C_{44} > 0$, $C_{55} > 0$, $C_{66} > 0$. Acoustic phonon modes significantly influence thermal transport processes in semiconductors. Using the calculated elastic properties in Tables S1 and S2, the lattice thermal conductivity $\kappa$ can be estimated via the Slack model[64], defined as $\kappa = A \frac{\bar{M}\delta n^{1/3}\Theta^3}{\gamma^2 T}$. In this model, $\bar{M}$ represents the average atomic mass, $\delta$ is the cube root of the average atomic volume, $n$ is the number of atoms in the primitive cell, determining the number of phonon branches, $\Theta$ denotes the acoustic Debye temperature, and $\gamma$ is the Grüneisen parameter. The coefficient $A$ in the Slack model[65,66] is commonly calculated according to $A = \frac{2.436 \times 10^{-6}}{1 - \frac{0.514}{\gamma} + \frac{0.228}{\gamma^2}}$. The estimated lattice thermal conductivity using the Slack model is 1.26 W/mK at 300 K, comparable to that of high-performance thermoelectric materials such as Bi$_2$Te$_3$[67], PbTe[6], and GeTe[68]. While the Slack model can quickly predict lattice thermal conductivity, it tends to underestimate these values[69]. Therefore, calculating lattice thermal conductivity using three-phonon and four-phonon scattering processes provides a more accurate prediction closer to experimental results. This aspect will be discussed in Section 3.3.

Figs. 1(a) and 1(b) present the electronic band structure and density of states (DOS) of $Mg_2GeSe_4$, respectively, calculated using the HSE06 hybrid functional including SOC. The conduction band minimum (CBM) and the valence band maximum (VBM) are located between the $\Gamma$ and X points and at the $\Gamma$ point, respectively. $Mg_2GeSe_4$ exhibits a direct band gap of 1.42 eV and an indirect band gap of 1.28 eV, as indicated by gray and black arrows. The CBM forms a single valley between the $\Gamma$ and X points. The density of states at the VBM is significantly higher than that at the CBM, primarily due to Se p states, with minor contributions from Ge p and d states. The calculated band gap is slightly smaller than previous results (1.57 eV)[16] obtained via the CASTEP package[70] using the LDA functional. Experimental measurements by diffuse reflectance spectroscopy reported a band gap of 2.02 eV which, as is typical due to photon absorption mechanisms, is slightly higher than the true electronic band gap[71,72]. Unfortunately, no angle-resolved photoelectron spectroscopy (ARPES) data is available for $Mg_2GeSe_4$, limiting direct experimental comparisons.

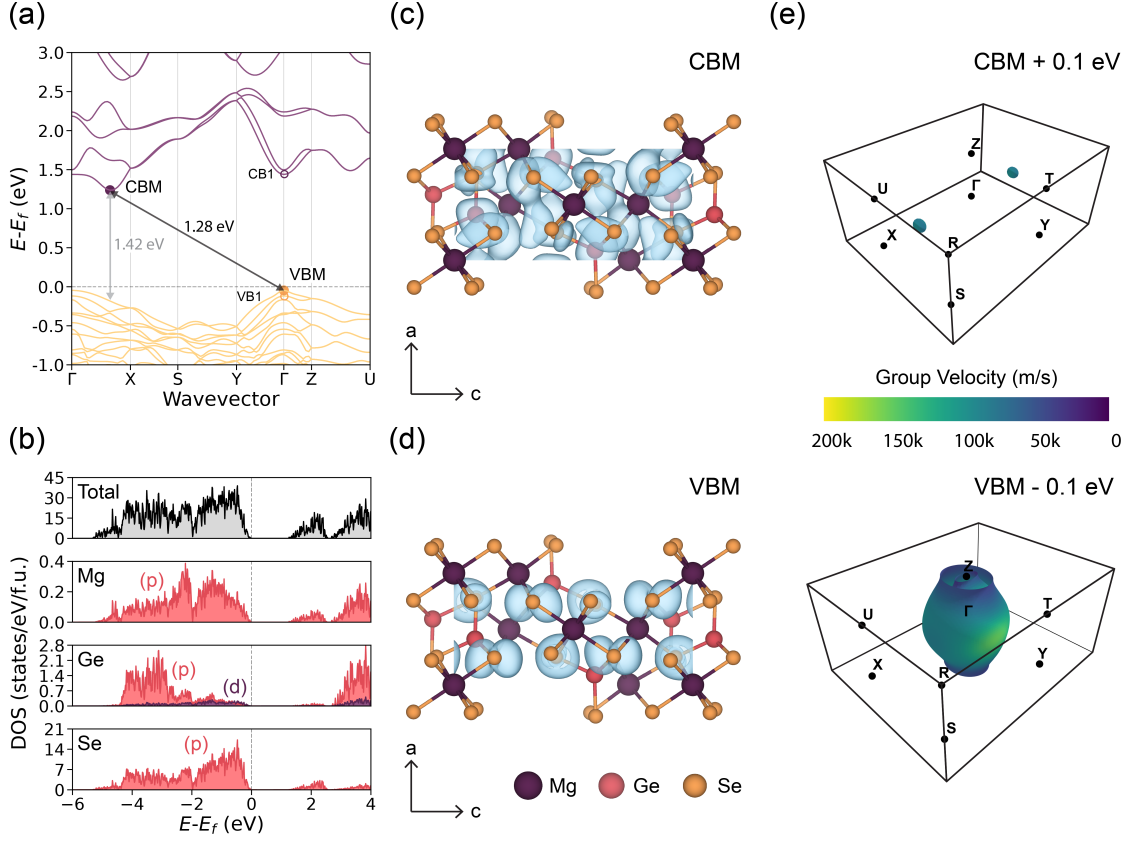

**Fig. 1.** (a) Electronic band structure and (b) density of states calculated by HSE06 + SOC functionals. The band decomposed charge densities at (c) CBM and (d) VBM. The isosurface is set to be 0.0005 bohr$^{-3}$. (e) The Fermi surface plotted at 0.1 eV above CBM and below VBM with IFermi code[73]. The color bar represents the magnitude of carrier group velocities.

Band-decomposed charge densities at the CBM and VBM are illustrated in Figs. 1 (c) and 1 (d). At the CBM, the electron cloud is relatively delocalized. In contrast, the charge density at the VBM is more localized, which hinders charge transport in the case of hole doping. This explains why the electronic transport properties with electron doping surpass those for hole doping.

Effective masses (Table S3) were calculated using the equation, $\frac{1}{m^*} = \frac{\partial^2 E(\mathbf{k})}{\partial \mathbf{k}^2} \frac{1}{\hbar^2}$, where $E(\mathbf{k})$ represents the band energy as a function of the electron wave vector $\mathbf{k}$. The CBM shows a higher electronic band dispersion than the VBM, resulting in lower effective masses and thus higher carrier mobility. To enhance the Seebeck coefficient, the Mott formula[4,74], $S = \frac{\pi^2}{3}\frac{k_B^2 T}{e}\left[\frac{DOS(E)}{n(E)} + \frac{1}{\mu(E)}\frac{d\mu(E)}{dE}\right]\bigg|_{E=E_F}$, is often invoked in connection with band engineering, where $e$ is the elementary charge, and $DOS(E)$, $n(E)$, and $\mu(E)$ represent the energy-dependent density of states, carrier concentration, and mobility, respectively. Band degeneracy is one strategy used in band engineering to increase the density of states near the Fermi level, although multi-band features can produce similar effects. In Fig. 1 (a), an energy difference of 0.2 eV is seen between the two lowest conduction bands ($E_{CB1} - E_{CBM}$), while only 0.07 eV separate the two highest valence bands ($E_{VBM} - E_{VB1}$). As shown in Fig. 1(e), two electron pockets are located between the Γ and X points for electron doping, and two hole pockets with high anisotropy at the center of the Brillouin zone for the hole doping. Due to the lower dispersion at the VBM and VB1, these two bands have higher effective masses. As a result, hole doping would yield a higher Seebeck coefficient than electron doping.

3.2 Electronic Transport Properties

Fig. 2 (a) presents the Seebeck coefficients $|S|$ of $Mg_2GeSe_4$ at various temperatures as a function of carrier concentration for both n-type and p-type doping. Like most thermoelectric materials, Seebeck coefficient at a given temperature decreases as carrier concentration increases. Unlike certain thermoelectric materials, however, they exhibit a bipolar effect induced by a small bandgap, causing the Seebeck coefficient to decrease at high temperatures and low carrier concentrations due to the cancellation of positive and negative Seebeck values[75,76]. Fig. 2 (b) shows the electrical conductivity $\sigma$ of $Mg_2GeSe_4$ at various temperatures as a function of carrier concentration for n-type and p-type doping. The electrical conductivity can be given by $\sigma = ne\mu$, where $n$ is the carrier concentration, $e$ is the elementary charge, $\mu = e\tau/m^*$ is the mobility, and $\tau$ is the relaxation time. The scattering rate $\tau_{ep}^{-1}$, has a significant impact on mobility and electrical conductivity, along with the effective mass. Key factors include acoustic deformation potential (ADP), ionized impurity (IMP), polar optical phonon (POP), and piezoelectric (PIE) scattering. As shown in Fig. S3, n-type $Mg_2GeSe_4$ demonstrates higher mobility than p-type, mainly due to its smaller band effective mass. Notably, polar optical phonon scattering is a major contributor to electron mobility. These findings align with previous studies on other polar semiconductors, such as the high-performance thermoelectric materials $Na_2TlSb$[77] and $Cs_3Cu_2I_5$[78]. Furthermore, the effect of piezoelectric scattering on mobility is minimal, suggesting it has a negligible impact.

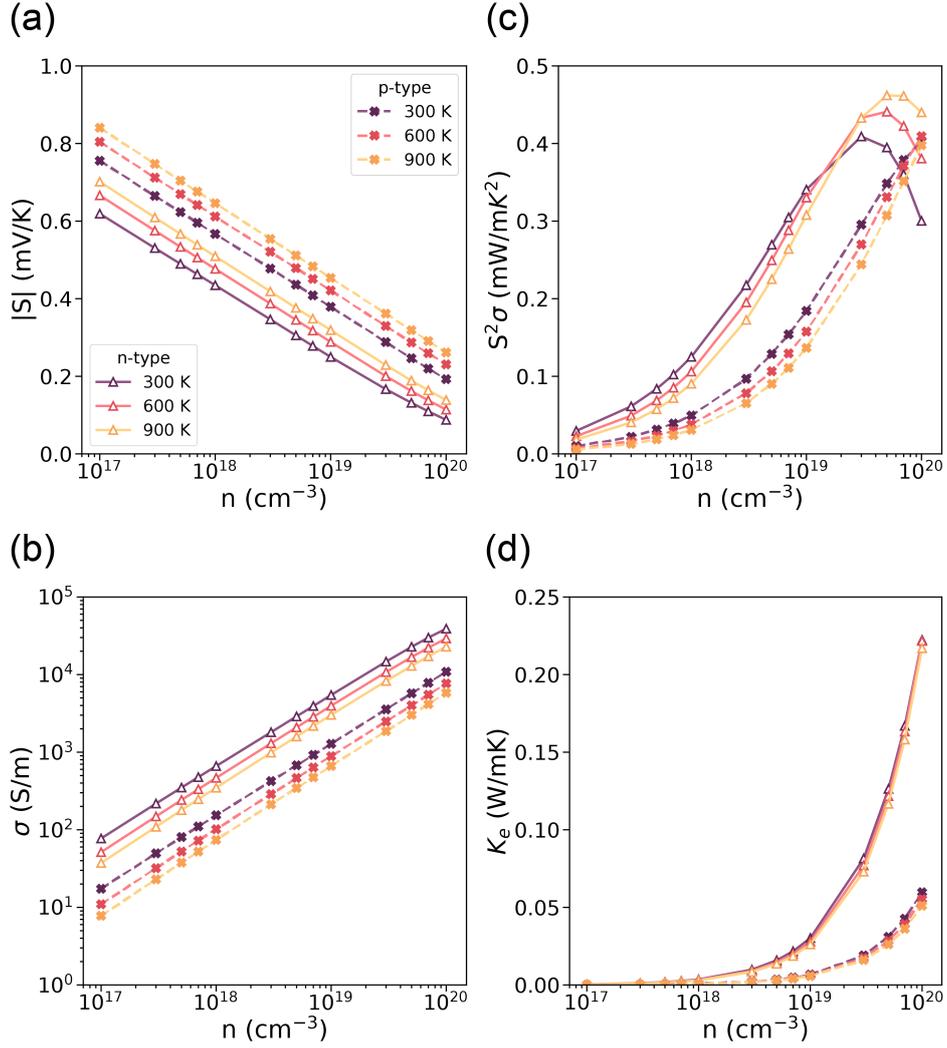

**Fig. 2.** The calculated (a) Seebeck coefficient *S*, (b) electrical conductivity $\sigma$, (c) power factor $S^2\sigma$, and (d) electronic thermal conductivity $\kappa_e$ as a function of carrier concentration for n-type and p-type.

Due to the opposing dependence of Seebeck coefficient and electrical conductivity on carrier concentration, we further analyze the power factor $PF = S^2\sigma$ to evaluate the electronic transport properties. Fig. 2 (c) shows the power factor at various temperatures as a function of carrier concentration for both n-type and p-type doping. As carrier concentration increases, the power factor initially rises and then declines. The power factor for n-type doping is higher than that for p-type doping, especially at high carrier concentrations.

Additionally, the trend in electronic thermal conductivity $\kappa_e$, parallels that of electrical conductivity in keeping with their linear relationship. The $\kappa_e$ for n-type and p-type doping are calculated according to the Wiedemann–Franz law[79], $\kappa_e = L\sigma T$, where $L$ denotes the Lorenz number. Fig. 2 (d) presents electronic thermal conductivity as a function of carrier concentration at various temperatures. Clearly, the electronic thermal conductivity shows a progressively higher contribution to total thermal conductivity as carrier concentration increases.

3.3 Phonon Transport Properties

Although the MLIP (MACE model) demonstrates excellent RMSE of 0.4 meV/atom for predicted energies and 19.5 meV/Å for force tensors on the test dataset, we first compared the MLIP and DFT predictions for the second-order interatomic force constants (IFC), phonon dispersions, and the corresponding partial density of states (PDOS) using the Phonopy code[80,81] to accurately predict phonon transport properties. Fig. S4 (a) shows strong agreement in phonon dispersion and PDOS from MLIP predictions. The phonon dispersion shows no imaginary frequencies across the Brillouin zone, indicating the dynamical stability of $Mg_2GeSe_4$ at 0 K. The second-order IFC matrices from DFT and MLIP were further compared using the Frobenius norm, with $\Phi_{Accuracy} = 1 - \frac{\|\Phi_{MLIP}-\Phi_{DFT}\|_2}{\|\Phi_{DFT}\|^2} \cdot 100\%$. High accuracy of 98.71% and 99.73% was achieved for the second-order IFCs and phonon frequencies, respectively. In Fig. S2 (b), we compared the calculated phonon dispersion with and without the inclusion of non-analytic corrections (NAC). The NAC was applied to the harmonic dynamical matrix to accurately capture the longitudinal optical–transverse optical (LO–TO) splitting near the Γ point.

The MLIP are employed to calculate temperature-dependent renormalized phonon dispersion and corresponding PDOS using the SCP method, as shown in Fig. 3 (a). Notably, no negative phonon frequencies appear within the first Brillouin zone, indicating the dynamical stability of $Mg_2GeSe_4$ across the entire temperature range. The three acoustic phonon modes exhibit several low-frequency phonons (< 2 THz) along the high-symmetry Γ–X, Γ–Y, and Γ–Z paths. Based on the PDOS shown in Fig. 3 (a), the Se atom plays a dominant role in vibrational modes. The frequency range between 5.5 THz and 7 THz is primarily contributed by Mg, while above 7 THz, Ge and Se contribute similarly. Acoustic phonon modes exhibit limited temperature sensitivity, while optical phonon modes, especially high-frequency optical phonons, undergo significant hardening with increasing temperature. Although the SCP method renormalizes phonon frequencies and thereby alters the heat capacity, it shows no significant difference when compared with heat capacity calculations based on the harmonic approximation (HA), as illustrated in Fig. 3 (b).

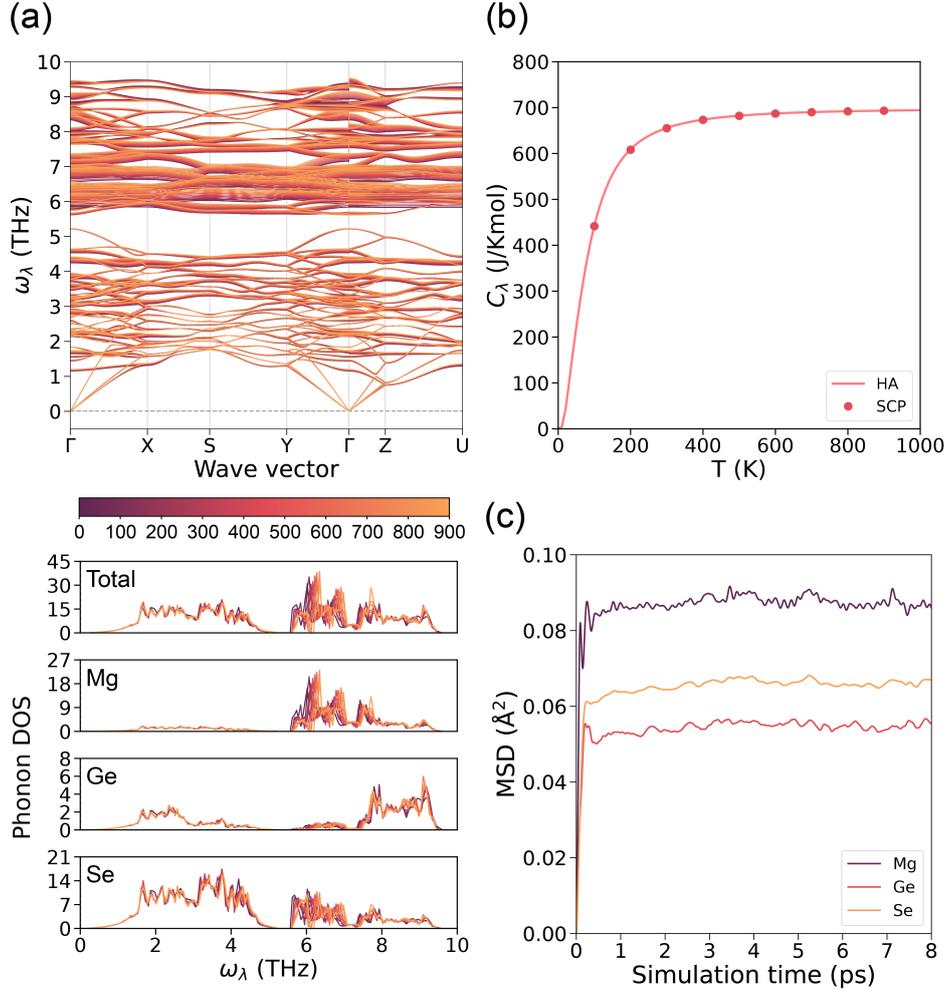

**Fig. 3.** Calculated temperature-dependence of (a) phonon dispersion and phonon density of states, and (b) heat capacity. (c) Mean square displacement (MSD) obtained via molecular dynamics at 300 K.

The mean square displacement (MSD) of atomic vibrations is a key parameter for assessing atomic displacements from their equilibrium positions. High MSD values indicate weak bonding interactions. Fig. 3 (c) shows the MSDs of different atoms in $Mg_2GeSe_4$ at 300 K. Relatively higher MSDs of Mg and Se atoms, compared to Ge atoms, imply weaker chemical bonding for Mg-Se bonding. The corresponding radial distribution functions (RDFs) are shown in Fig. S5, displaying well-defined peaks that broaden and shift to larger radial distances with increasing temperature, indicating more pronounced thermal vibrations.

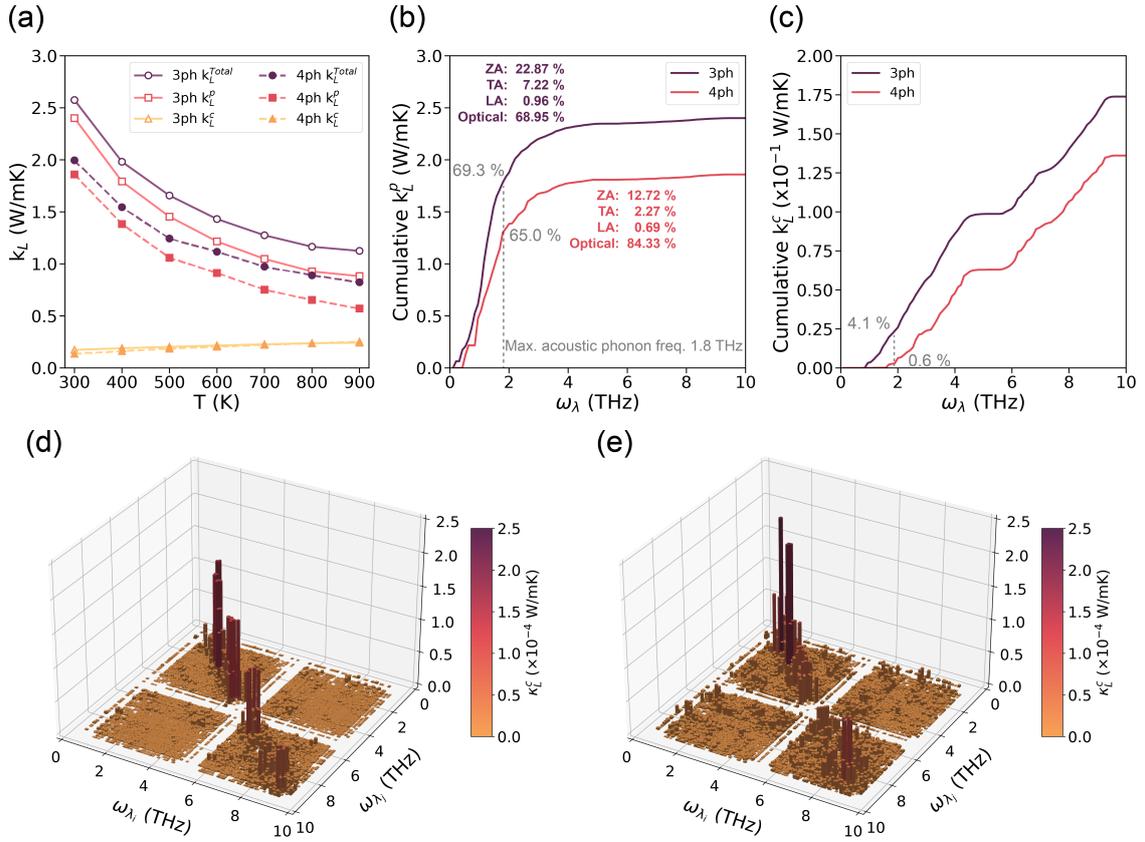

**Fig. 4.** (a) The calculated temperature-dependent three and four phonon lattice thermal conductivity. Cumulative (b) particle-like and (c) wave-like contributions to thermal conductivity as a function of frequency. The three-dimensional visualization $\kappa_L^c(\omega_{\lambda_i}, \omega_{\lambda_j})$ of the contributions to the wave-like thermal conductivity at (d) 300 K and (e) 900 K.

Fig. 4 (a) shows the total $\kappa_L^{Total}$, particle-like $\kappa_L^p$, and wave-like $\kappa_L^c$ lattice thermal conductivity for Mg$_2$GeSe$_4$, calculated considering three-phonon and four-phonon interactions, as a function of temperature. The results exhibit a strong temperature dependence. Including four-phonon scattering, the calculated lattice thermal conductivity decreases from approximately 2.57 to 1.99 W/mK at 300 K, representing a reduction of 22.5 %. At 900K, the reduction reaches 26.7 %. Furthermore, the $\kappa_L^p$ decreases as the temperature rises, while $\kappa_L^c$ exhibits the opposite trend.

Fig. 4 (b) and (c) show the particle-like and wave-like cumulative contributions to the lattice thermal conductivity as a function of frequency at 300 K for Mg$_2$GeSe$_4$. In Fig. 4 (b), the contribution by particle-like three-phonon and four-phonon interactions up to the maximum acoustic phonon frequency (1.8 THz) amounts to 69.3 % and 65.0 %, respectively. However, the contribution from acoustic phonon modes (ZA, TA, and LA) is 31.05 % for the three-phonon and 15.67 % for the three- and four-phonon combined, as shown in the inset of Fig. 4 (b). This indicates that scattering from acoustic and low-frequency optical modes is quite significant. Furthermore, the wave-like lattice thermal conductivity from three-phonon and four-phonon interactions exceeds 95.9 % and 99.4 % above 1.8 THz, respectively, which is attributed to the presence of high-frequency flat optical phonon bands. Fig. 4 (d) and (e) present three-dimensional visualizations $\kappa_L^c\left(\omega_{\lambda_i}, \omega_{\lambda_j}\right)$ of the frequency-dependent contributions to wave-like thermal conductivity at 300 K and 900 K. The phonon pairs with substantial contributions to $\kappa_L^c$ are primarily along the diagonal at both temperatures, although off-diagonal phonon pair contributions become more pronounced at 900 K.

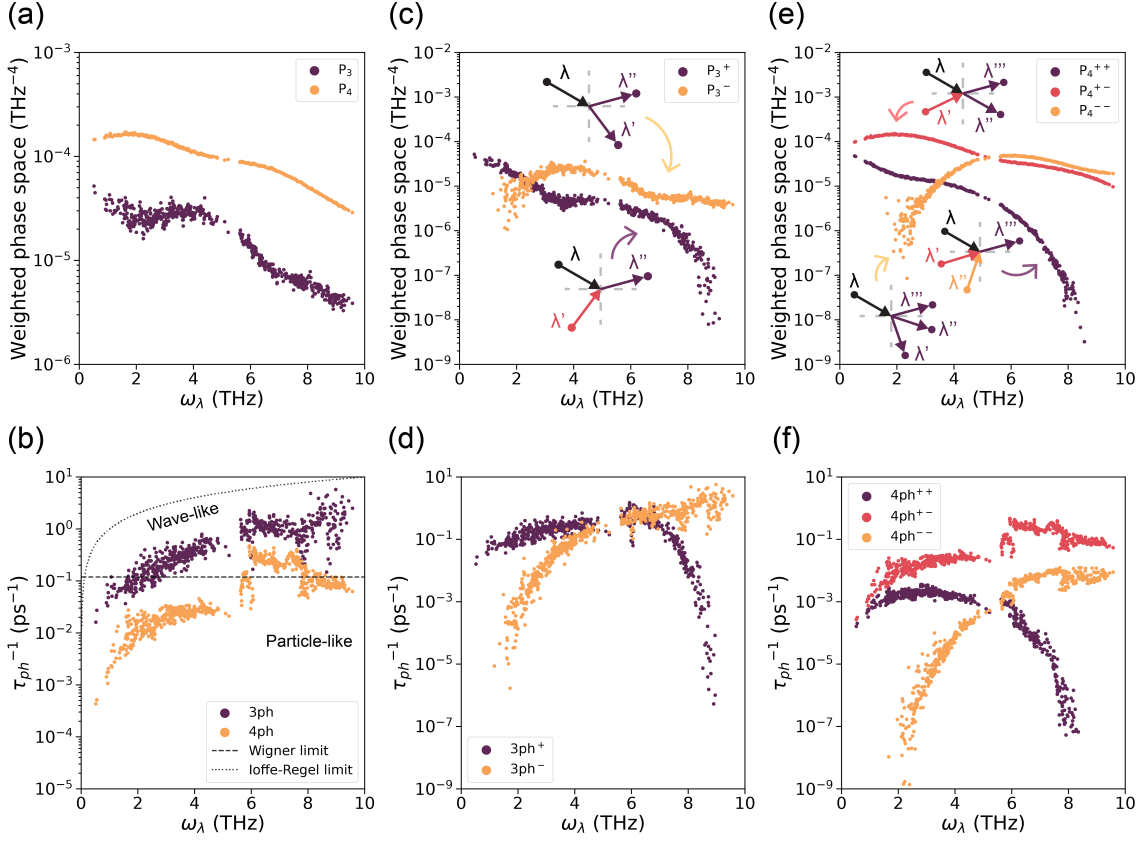

**Fig. 5.** (a) The total weighted scattering phase space as a function of frequency, including three-phonon ($P_3$) and four-phonon ($P_4$) scattering phase space. (b) The different scattering rates as a function of frequency, including three-phonon (3ph), four-phonon (4ph) scattering. (c) The weighted scattering phase and (d) the scattering rates under 3ph mechanism as a function of phonon frequency, including absorption ($\lambda + \lambda' \rightarrow \lambda''$) and emission ($\lambda \rightarrow \lambda' + \lambda''$) processes. (e) The weighted scattering phase and (f) the scattering rates under 4ph mechanism as a function of frequency, including combination ($\lambda + \lambda' + \lambda'' \rightarrow \lambda'''$), redistribution ($\lambda + \lambda' \rightarrow \lambda'' + \lambda'''$), and splitting ($\lambda \rightarrow \lambda' + \lambda'' + \lambda'''$).

To gain insight into phonon scattering mechanisms, we consider the phonon phase space, as shown in Fig. 5. The four-phonon scattering phase space ($P_4$) values are generally higher than those for the three-phonon scattering case ($P_3$) across the phonon frequency spectrum of $Mg_2GeSe_4$, as shown in Fig. 5 (a). Additionally, we evaluated the phonon relaxation time by incorporating multiple scattering mechanisms based on Matthiessen's rule: $\tau_{ph}^{-1} = \tau_{3ph}^{-1} + \tau_{4ph}^{-1}$. Here, $\tau_{3ph}$ and $\tau_{4ph}$ denote phonon relaxation times due to three and four phonon scatterings, respectively. Fig. 5 (b) illustrates the scattering rates of $Mg_2GeSe_4$ across different scattering mechanisms as a function of phonon frequency. Fig. 5 (c) displays a phase-space decomposition of the three-phonon scattering processes as a function of frequency, with purple markers representing absorption processes ($\lambda + \lambda' \rightarrow \lambda''$) and yellow markers indicating emission processes ($\lambda \rightarrow \lambda' + \lambda''$). In other words, low-frequency phonons predominantly engage in absorption processes to produce high-frequency phonons, whereas high-frequency phonons are mainly generated through their three-phonon scattering processes, as illustrated in Fig. 5 (d). This behavior aligns with the principles of energy and momentum conservation. The inset of Fig. 5 (c) provides a detailed breakdown of the three-phonon scattering channels. Additionally, Fig. 5 (e) shows the phase-space decomposition for four-phonon scattering processes, which involve three types of scatterings: combination ($\lambda + \lambda' + \lambda'' \rightarrow \lambda'''$), redistribution ($\lambda + \lambda' \rightarrow \lambda'' + \lambda'''$), and splitting ($\lambda \rightarrow \lambda' + \lambda'' + \lambda'''$), represented by purple, red, and yellow markers, respectively. For the four-phonon scattering process, the phase space is predominantly governed by redistribution processes. The redistribution process aligns more effectively with the selection rules, resulting in significantly higher scattering rates, as shown in Fig. 5 (f). This highlights the critical role of redistribution process in contributing to the overall scattering rates, especially in the context of four-phonon interactions.

3.4 Thermoelectric Performance

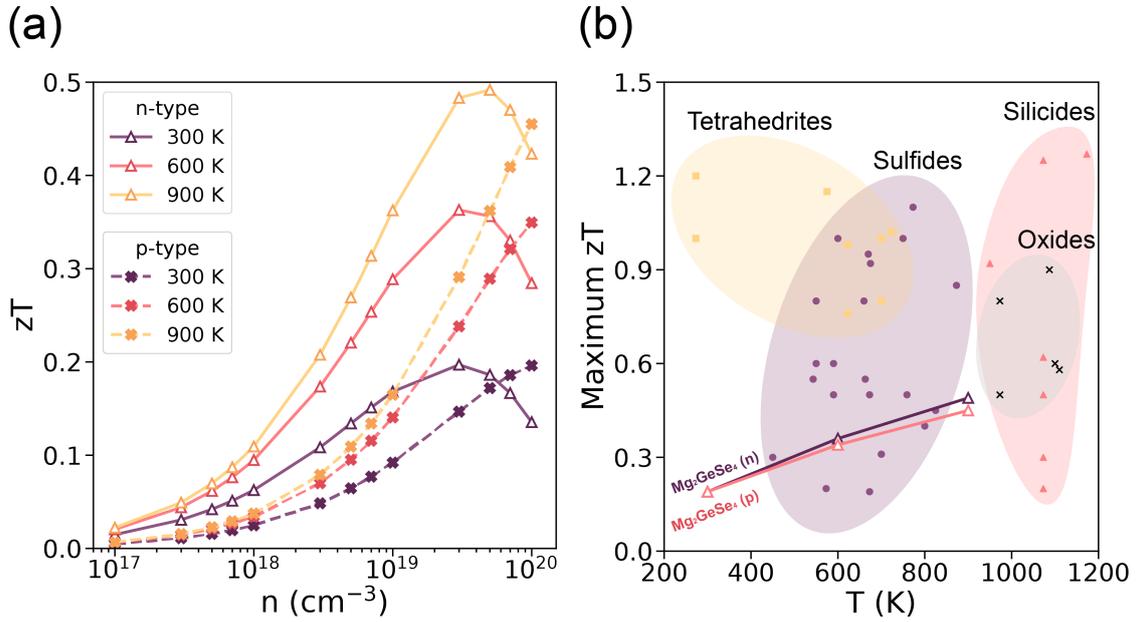

**Fig. 6.** (a) The calculated figure of merit zT at T = 300 K, 600 K, and 900 K as a function of carrier concentration for n-type and p-type. (b) Comparison of maximum ZT as a function of temperature for $Mg_2GeSe_4$ with respect to eco-friendly thermoelectric materials including sulfides[82–85], silicides[86–92], tetrahedrites[93–97], and oxides[98–104].

Following a comprehensive assessment of the thermal and electrical transport parameters, we have calculated the figure of merit zT, for both n-type and p-type $Mg_2GeSe_4$. Fig. 6 (a) displays the n-type zT values across different temperatures as a function of carrier concentration. Optimal n-type zT values were found to be 0.19, 0.36, and 0.49 at 300 K, 600 K, and 900 K, respectively, at carrier concentrations of $3 \times 10^{19}$, $3 \times 10^{19}$, and $5 \times 10^{19}$ cm$^{-3}$. Similarly, the p-type zT values are shown as a function of carrier concentration, with optimal values of 0.19, 0.34, and 0.45 at 300 K, 600 K, and 900 K, respectively, at a carrier concentration of $10^{20}$ cm$^{-3}$. As shown in Fig. 6 (b), these findings suggest that both n-type and p-type Mg$_2$GeSe$_4$ exhibit promising thermoelectric performance. With further performance enhancements through element doping, Mg$_2$GeSe$_4$ shows significant potential as a high-performance thermoelectric material.

## 4. CONCLUSION

We explore thermal and electrical transport in chalcogenide Mg$_2$GeSe$_4$ within a SCP theoretical framework to obtain temperature-dependent renormalized phonon frequencies, which are found to show high-frequency optical phonon hardening, underscoring substantial effects of lattice anharmonicity. Utilizing a message-passing neural network architecture, the MACE model was trained to achieve DFT-level precision in energy and atomic force predictions, resulting in accurate phonon transport prediction for particle-like and wave-like lattice thermal conductivity. Our analysis reveals that four-phonon interactions and off-diagonal heat flux components have a considerable impact on the thermal transport properties of Mg$_2$GeSe$_4$. Four-phonon processes increase scattering rates and lead to reduction in thermal conductivity. An examination of the electronic band structure indicates that Mg$_2$GeSe$_4$ should host a substantial Seebeck coefficient driven by the presence of multi-band features in the valence bands near the Fermi level. Our theoretical estimates of the maximum figure of merit zT of 0.49 for the n-type and 0.45 for the p-type Mg$_2$GeSe$_4$, suggest that Mg$_2$GeSe$_4$ is a promising candidate for thermoelectric applications. Our study gives insight into how various scattering processes contribute to the thermal and electrical transport in Mg$_2$GeSe$_4$ and provide a comprehensive machine-learning based scheme for their accurate and efficient calculation as a tool for rational design and optimization of high-performing thermoelectric materials.

**SUPPORTING INFORMATION**

Supporting Information is available from the Wiley Online Library or from the author.

**ACKNOWLEDGEMENTS**

HL acknowledges the support by the National Science and Technology Council (NSTC) in Taiwan under grant number MOST111-2112-M-001-057-MY3. The work at Northeastern University was supported by the National Science Foundation through NSF-ExpandQISE award NSF-OMA-2329067 and benefited from the resources of Northeastern University's Advanced Scientific Computation Center, the Discovery Cluster, and the Massachusetts Technology Collaborative award MTC-22032.

**CONFLICT OF INTERESTS**

The authors declare no competing interests.

**AUTHOR CONTRIBUTIONS**

**Hao-Jen You** developed the research methodology, performed DFT calculations, collected data, and trained the machine learning model with assistance from **Yi-Ting Chiang**. **Hao-Jen You** analyzed the results and collaborated with **Arun Bansil** and **Hsin Lin** to validate the findings and write the paper.

**DATA AVAILABILITY STATEMENT**

The training and testing data for the trained $Mg_2GeSe_4$ MLIP is freely available at https://github.com/Youhaojen/supplementary_information.